# α-decay half-lives for Pb isotopes within Gamow-like model

**Dashty T. Akrawy**[a,b*]

[a] *Akre Coputer Institute, Ministry of Education, Akre, Kurdistan, Iraq.*
[b] *Becquereal Institute For Radiation Research and Measurements, Erbil, Kurdistan, Iraq*
[*]akrawy85@gmail.com

**Abstract**

We studed alpha decay half-lives of Pb isotopes in the range $178 \leq A \geq 220$, whithin Gamow-like model (GLM) which is based on Gamow theory. The empirical formulae like Royer formula, Universal Decay Low (UDL), Viola-Seaborg formula (VSS), Semi-empirical formula for Poenaru (SemFIS) and Denisov & Khudenko formula (DEKH) are used to calculate alpha decay half-lives. The results are compared with experimental data, and also with other theoretical models, the results are in a good agreement was achieved with experimental data.

**Keywords**: α-decay; Gamow-like model; Semi-empirical formula

**PACS:** 23.60. +e

## 1. Introduction

The study of alpha decay is one of the important modes to describe nuclear structure [1], it can be successfuly described by qunatum theory [2]. The alpha decay was described in 1928 [3] in base of quantum tunneling. A simple formula for α-decay half-lives was derived using Wentzel–Kramers–Brillouin (WKB) approximation theory by A.Zdeb etal [4] for penetration of the Coulomb barrier with a square well for the nuclear part. GLM depended only one adjustable parameter, the radius constant, which is possible to reproduce with a good accuracy all existing data for decays of even-even nuclei. There are more model that was used to determine alpha decay half-lives such as generalized the liquid drop model (GLDM) which are used by Royer to describe α-decay half-lives [5]. By considering Yukawa proximity potential (YPP) from ground-state to ground-state, Javadimanesh et.al was used to calculate α-decay half-lives [6]. Preformed cluster model (PCM) was used by Kumar, to evaluate alpha decay half-lives for superheavy nuclei [7]. Double-folding with the standard M3Y nucleon-nucleon interaction (M3YDDIM) was used by Ren to calculate alpha decay half-lives [8]. New empirical formula was used to calculate alpha decay half-lives by Akrawy, which is modified the Royer formula [9]. Akrawy used (GLM) to calculate α-decay half-lives for Lv isotopes [10]. In this study we calculated the alpha decay decay half-lives for Pb isotopes using Gamow-Like Model (GLM), we compared the results with CPPM theoretical model [11] and with experimental data. Also we used some empirical formula to calculate alpha decay half-lives and to compare with GLD, like Royer formula [12], Universal Decay Low (UDL) [13], Viola-Seaborg formula (VSS) [14], Semi-empirical formula for Poenaru et. Al. (SemFIS) [15], and finally with Denisov & Khudenko formula (DEKH) [16].

## 2. Material Methods

A.Zdeb et. al [4], developed and used a simple model to calculate alpha decay half-lives, this formula baised on Gamow theory. this model formula is contain Coulomb potential adjusted parameter.

The penetration probability of tunneling alpha particle through barrier potential are calculated within Wentzel-Kramers-Brillouin (WKB) approximation using the following integral [4,17]

$$P = \exp\left[-\frac{2}{\hbar}\int_R^b \sqrt{2\mu(V(r)-Q)}\,dr\right] \qquad (1)$$

where μ is reduced mass, $Q$ is energy relased. $a$ and $b$ are tunniling point of integral which are given as

$$R = r_0\left(A_1^{1/3} + A_2^{1/3}\right) \qquad (2)$$

where $R$ is the spherical square well radius, $A_1$ and $A_2$ are atomic number of parent and daughter nuclei. $r_0$ value has been taken from Ref. [5],

and

$$b = \frac{Z_1 Z_2 e^2}{Q} \qquad (3)$$

where $e^2$ is the electron charge [18], $Z_1$ and $Z_2$ are the atomic number of parent and daughter nuclei respectively.

The potential energy $V(r)$ is given by the following conditions

$$V(r) = \begin{cases} -V_0 & 0 \le r \le R \\ \dfrac{Z_1 Z_2 e^2}{r} & r > R \end{cases} \qquad (4)$$

Where $V_o$ is the depth of the potential well which is equal to $V_0 = 25A_1$ [19]. The α-decay half-life can be evaluate by using [20]

$$T_{1/2} = \frac{\ln 2}{vP} \qquad (4)$$

where $v$ is the assault frequency which is equal to $10^{20}$ s$^{-1}$ [21].

1- **Royer empirical formula (Royer)**

Royer predicted new formula for alpha decay half-lives [12,22-23] by fitting 373 nuclei which have transition from ground-state to ground state.

$$\log_{10}[T_{1/2}(s)] = a - bA^{1/6}\sqrt{Z} + \frac{cZ}{\sqrt{Q}} \qquad (5)$$

where $a$, $b$ and $c$ are coefficinet are determined by least squre fitting method, $A$ and $Z$ are mass number of and atomic number of parent nuclei respectively, and $Q$ represented energy released during the reaction.

**Universal decay law formula (UDL)**

Qi et al, was predicted the universal decay law for cluster decay from micrscopic mechanism process of the radioactive decay [13], the relation is found to be as [24,25]

$$\log_{10}[T_{1/2}(s)] = aZ_c Z_d \sqrt{\frac{A}{Q_c}} + b\sqrt{AZ_c Z_d \left(A_d^{1/3} + A_c^{1/3}\right)} + c \quad (6)$$

$$= a\chi' + b\rho' + c \quad (7)$$

where $A = \sqrt{\frac{A_d A_c}{(A_d + A_c)}}$, and the $a$, $b$ and $c$ are constants which are used was determined by fitting to the experimental data of α-decay and cluster [26]. $b\rho' + c$ is the effects term that include the cluster in the parent nucleus.

2- **The Viola-Seaborg semi-empirical formula (VSS)**

From the Geiger-Nuttall law, the VSS formula was determined by Sobiczewski et al. [27], is given as

$$\log_{10}[T_{1/2}(s)] = \frac{(aZ+b)}{\sqrt{Q}} + cZ + d + h_{\log} \quad (8)$$

Here $Z$ is the parent atomic number, $Q$ is the energy released during the reaction in unit MeV, half-life in seconds, the coefficients $a$, $b$, $c$, and $d$ are adaptable parameters and the parameter $h_{\log}$ evidence the hindrance factor related with odd proton and odd neutron numbers [28], as given by Viola et al. [29]. More recent value was determined by Tiekuang Dong et. Al [14].

3- **Semi-empirical formula for Poenaru et. Al. (SemFIS)**

Poenaru et al. modified semi-empirical formula for α-decay half-life which is based on fission theory for α-decay yield which is given as [15]

$$\log_{10}[T_{1/2}(s)] = 0.43429 K_s \chi - 20.446 \quad (9)$$

where

$$K_s = 2.52956 Z_d \left[\frac{A_d}{AQ_\alpha}\right]^{1/2} \left[\arccos\sqrt{r} - \sqrt{r(1-r)}\right] \quad (10)$$

and

$$r = 0.423 Q_\alpha \frac{\left(1.5874 + A_d^{1/3}\right)}{Z_d} \quad (11)$$

and the numerical coefficient $\chi$, close to the unity, is a second order polynomial

$$\chi = B_1 + B_2 y + B_3 z + B_4 y^2 + B_5 yz + B_6 z^2 \quad (12)$$

The reduced variable $y$ and $z$, expressed the distance from the closest magic-plus-one neutron and proton numbers $N_i$ and $Z_i$ is given as

$$z = \frac{(N - N_i)}{(N_{i+1} - N_i)}, \qquad N_i < N \le N_{i+1} \qquad (13)$$

$$y = \frac{(Z - Z_i)}{(Z_{i+1} - Z_i)}, \qquad Z_i < Z \le Z_{i+1} \qquad (14)$$

with $N_i$ = …, 51, 83, 127, 185, 229, …, and $Z_i$ = …, 29, 51, 83, 127, ….; Bi have been obtained by using a high quality selected data set of α-decay.

### 4- Denisov & Khudenko formula (DEKH)

From the Guy Royer empirical formula [23], Denisov & Khudenko [16,30] developed empirical formula for α-decay half-life between ground state to ground state α-transition of parent and daughter nuclei, the formula is evaluated of α-decay half-lives for even-even, even-odd, odd-even and odd-odd nuclei:

$$\log_{10}[T_{1/2}(s)] = -a - \frac{bA^{1/6}\sqrt{Z}}{\mu} + \frac{cZ}{\sqrt{Q_\alpha}} + \frac{d\sqrt{\ell(\ell+1)}}{QA^{-1/6}} - e\left((-1)^\ell - 1\right) \qquad (15)$$

Here A and Z are the mass number and atomic number of parent nucleus, respectively, $\ell$ is the orbital moment of emitted α particle, and $\mu = (A/(A-4))^{1/6}$. $Q$ is the reaction energy value. The α-particle emission from nuclei obeys the spin-parity selection rule [31,32]

$$\ell_{min} = \begin{cases} \Delta j & \text{for even } \Delta j \text{ and } \pi_p = \pi_d \\ \Delta j + 1 & \text{for odd } \Delta j \text{ and } \pi_p = \pi_d \\ \Delta j & \text{for odd } \Delta j \text{ and } \pi_p \ne \pi_d \\ \Delta j + 1 & \text{for even } \Delta j \text{ and } \pi_p \ne \pi_d \end{cases} \qquad (16)$$

where $\Delta j = |j_p - j_d|, j_p, \pi_p, j_d, \pi_d$ are spin and parity values of the parent and daughter nuclei, respectively, and $\ell = \ell_{min}$, the *a, b, c, d,* and *e* are free parameters have been evaluated by Denisov et al.[30].

### 3. Results and Discussion

The theoretical α-decay half-lives of Pb isotopes, which have atomic number Z = 82 within the range A = 178-220 have been studied by using Gamow-Like Model. The *Q*-value was taken in Ref. [33] which is evaluated experimentaly are listed in Table I. The evaluation of α-decay half-lives within GLM, which are compared with experimental data [26,33] are given in Table I.

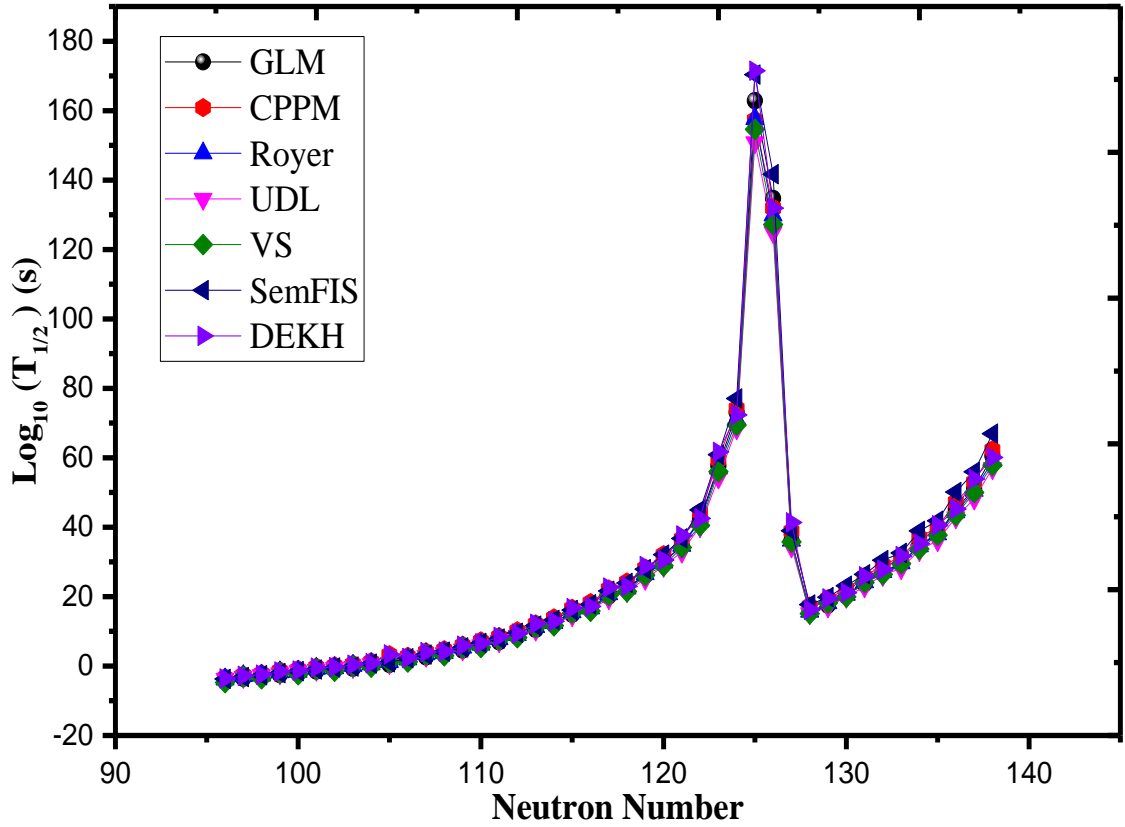

**Figure 1.** The comparison of the evaluated α-decay half-lives of Pb isotopes.

The semi-empirical formulae was also been done using the analytical Royer formula (Royer), the universal decay law (UDL), the Viola-Seaborg formula (VSS), the Semi-empirical formula for Poenaru et. Al. (SemFIS)], and finally analytical Denisov & Khudenko formula (DEKH) have also been evaluated and presented in Table I

The maximum value of logarithm ($T_{1/2}$) half-life is for the decay of the parent nuclei $^{207}$Pb, which have $N$=125, the maximum value of logarithm ($T_{1/2}$) half-life which is reference the behavior nearly doubly magic number, this observation indicate the role of neutron shell close in α-decay.

**Table 1.** The comparison of the evaluated α-decay half-lives of Pb isotopes with experimental data and with other theoretical formula in the range $178 \leq A \geq 220$, $Z = 82$.

| A | Q (MeV) | GLM | Exp. | CPPM | Royer | UDL | SemFIS | VSS | DEKH |
|---|---|---|---|---|---|---|---|---|---|
| 178 | 7.790 | -4.051 | -3.638 | -3.395 | -3.809 | -3.731 | -4.724 | -3.737 | -3.616 |
| 179 | 7.598 | -3.507 | -2.409 | -2.587 | -2.877 | -3.186 | -3.083 | -3.208 | -2.790 |
| 180 | 7.419 | -2.979 | -2.377 | -2.221 | -2.705 | -2.660 | -3.593 | -2.630 | -2.496 |
| 181 | 7.240 | -2.429 | -1.409 | -1.408 | -1.766 | -2.113 | -1.951 | -2.078 | -1.595 |
| 182 | 7.066 | -1.873 | -1.260 | -1.020 | -1.573 | -1.561 | -2.436 | -1.482 | -1.348 |
| 183 | 6.928 | -1.420 | -0.272 | -0.334 | -0.730 | -1.112 | -0.894 | -1.010 | -0.481 |
| 184 | 6.774 | -0.893 | -0.310 | 0.037 | -0.575 | -0.592 | -1.411 | -0.453 | -0.336 |

| A | Q (MeV) | GLM | Exp. | CPPM | Royer | UDL | SemFIS | VSS | DEKH |
|---|---|---|---|---|---|---|---|---|---|
| 185 | 6.695 | -0.625 | 0.799 | 0.520 | 0.081 | -0.326 | -0.057 | -0.151 | 0.393 |
| 186 | 6.470 | 0.207 | 1.083 | 1.217 | 0.540 | 0.490 | -0.271 | 0.705 | 0.796 |
| 187 | 6.393 | 0.490 | 2.185 | 3.217 | 1.214 | 0.769 | 1.096 | 1.036 | 3.356 |
| 188 | 6.109 | 1.638 | 2.185 | 2.750 | 1.984 | 1.889 | 1.191 | 2.207 | 2.261 |
| 189 | 5.870 | 2.671 | 2.431 | 4.037 | 3.421 | 2.892 | 3.299 | 3.310 | 3.984 |
| 190 | 5.697 | 3.456 | 3.591 | 4.679 | 3.810 | 3.654 | 3.028 | 4.111 | 4.112 |
| 191 | 5.460 | 4.604 | 4.250 | 5.938 | 5.364 | 4.763 | 5.243 | 5.347 | 5.809 |
| 192 | 5.221 | 5.846 | 5.902 | 7.210 | 6.195 | 5.958 | 5.414 | 6.610 | 6.532 |
| 193 | 5.010 | 7.019 | 6.551 | 8.410 | 7.776 | 7.083 | 7.645 | 7.886 | 8.377 |
| 194 | 4.738 | 8.657 | - | 10.152 | 8.983 | 8.649 | 8.195 | 9.556 | 9.359 |
| 195 | 4.450 | 10.563 | - | 12.097 | 11.295 | 10.464 | 11.130 | 11.594 | 12.122 |
| 196 | 4.226 | 12.181 | - | 13.818 | 12.456 | 11.999 | 11.648 | 13.252 | 12.882 |
| 197 | 3.889 | 14.896 | - | 16.606 | 15.565 | 14.566 | 15.351 | 16.131 | 16.666 |
| 198 | 3.709 | 16.492 | - | 18.243 | 16.679 | 16.072 | 15.838 | 17.782 | 17.165 |
| 199 | 3.343 | 20.164 | - | 22.164 | 20.725 | 19.520 | 20.441 | 21.655 | 22.590 |
| 200 | 3.151 | 22.340 | - | 24.201 | 22.374 | 21.559 | 21.472 | 23.925 | 22.940 |
| 201 | 2.857 | 26.111 | - | 28.182 | 26.517 | 25.082 | 26.150 | 27.915 | 28.827 |
| 202 | 2.590 | 30.090 | - | 32.017 | 29.879 | 28.790 | 28.884 | 32.079 | 30.552 |
| 203 | 2.335 | 34.525 | - | 36.609 | 34.674 | 32.911 | 34.172 | 36.761 | 37.617 |
| 204 | 1.969 | 42.375 | - | 44.243 | 41.711 | 40.183 | 40.545 | 44.988 | 42.552 |
| 205 | 1.468 | 57.638 | - | 59.408 | 56.947 | 54.271 | 56.008 | 60.877 | 61.670 |
| 206 | 1.135 | 73.027 | - | 74.146 | 71.068 | 68.436 | 69.415 | 77.063 | 72.323 |

**Table 1.** Continues.

| A | Q (MeV) | GLM | Exp. | CPPM | Royer | UDL | SemFIS | VSS | DEKH |
|---|---|---|---|---|---|---|---|---|---|
| 207 | 0.392 | 162.866 | - | 157.279 | 157.665 | 150.835 | 154.608 | 170.385 | 171.485 |
| 208 | 0.517 | 134.658 | - | 131.906 | 129.841 | 124.994 | 127.172 | 141.686 | 131.926 |
| 209 | 2.248 | 36.103 | - | 38.833 | 36.189 | 34.398 | 35.775 | 39.007 | 41.353 |
| 210 | 3.792 | 15.498 | - | 17.262 | 15.684 | 15.181 | 15.108 | 17.692 | 16.158 |
| 211 | 3.571 | 17.562 | - | 19.382 | 18.163 | 17.126 | 18.174 | 19.816 | 19.455 |
| 212 | 3.290 | 20.497 | - | 22.283 | 20.559 | 19.883 | 19.935 | 23.135 | 21.102 |
| 213 | 3.020 | 23.707 | - | 25.720 | 24.161 | 22.888 | 24.082 | 26.419 | 25.836 |
| 214 | 2.760 | 27.245 | - | 29.140 | 27.102 | 26.190 | 26.401 | 30.482 | 27.738 |

| 215 | 2.620 | 29.361 | - | 31.413 | 29.653 | 28.164 | 29.494 | 32.550 | 31.678 |
| 216 | 2.300 | 34.947 | - | 37.021 | 34.536 | 33.356 | 33.740 | 38.919 | 35.277 |
| 217 | 2.150 | 37.983 | - | 40.057 | 37.993 | 36.173 | 37.695 | 41.866 | 40.549 |
| 218 | 1.850 | 45.168 | - | 47.201 | 44.361 | 42.824 | 43.428 | 50.135 | 45.241 |
| 219 | 1.650 | 51.020 | - | 52.990 | 50.553 | 48.230 | 50.023 | 55.935 | 53.905 |
| 220 | 1.390 | 60.469 | - | 62.276 | 59.017 | 56.944 | 57.860 | 66.924 | 60.105 |

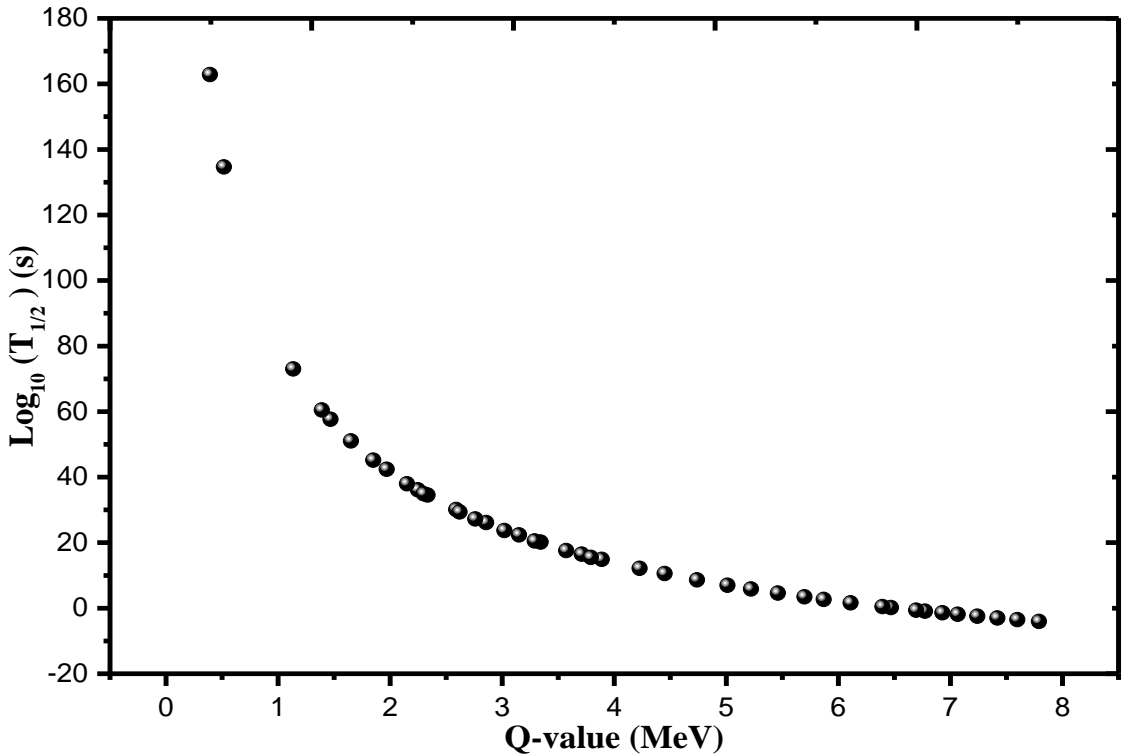

**Figure 2.** Correlation between Q-value vs. logarithm ($T_{1/2}$) half-lives of Pb isotopes.

From Table I, it can be seen clearly the GLM, CPPM, Royer, UDL, VSS, SemFIS and DEKH are the same orientation. It should be taken into account that GLM values match close to experimental data and Coulomb proximity potential model.

The correlation between $Q$-value and logarithm ($T_{1/2}$) α-decay half-lives was shown in Figure 2, it show the logarithm ($T_{1/2}$) α-decay half-lives exponential decrease with increasing $Q$-value

## 4- Conclusion

The α-decay half-lives for Pb isotopes within GLM have been performed. The comparison between GLM and experimental data have been done, also the our result was compared with CPPM and some empirical formulae such as Royer, UDL, VSS, SemFIS and DEKH, the result is match close to each other. The Q-value have the main role, which is decrease exponential with increasing logarithm ($T_{1/2}$) half-life. From the result we predict by GLM, the maximum value of logarithm ($T_{1/2}$) half-life of Pb isotopes is founded in $N = 125$, which is influence of the neutron shell close.

# Reference

[1] Shin, E., Lim, Y., Hyun, C. H., & Oh, Y. (2016). Nuclear isospin asymmetry in α decay of heavy nuclei. Physical Review C, 94(2), 024320.

[2] Hassanabadi, H., Javadimanesh, E., Zarrinkamar, S., & Rahimov, H. (2013). An angle-dependent potential and alpha-decay half-lives of deformed nuclei for 67≤ Z≤ 91. Chinese physics C, 37(4), 044101.

[3] Gamow, G. (1928). Zur quantentheorie des atomkernes. Zeitschrift für Physik, 51(3-4), 204-212.

[4] Zdeb, A., Warda, M., & Pomorski, K. (2013). Half-lives for α and cluster radioactivity within a Gamow-like model. Physical Review C, 87(2), 024308.

[5] Royer, G. (2000). Alpha emission and spontaneous fission through quasi-molecular shapes. Journal of Physics G: Nuclear and Particle Physics, 26(8), 1149.

[6] Javadimanesh, E., Hassanabadi, H., Rajabi, A. A., Rahimov, H., & Zarrinkamar, S. (2012). Alpha Decay Half-Lives of Some Nuclei from Ground State to Ground State with Yukawa Proximity Potential. Communications in Theoretical Physics, 58(1), 146.

[7] Kumar, S., Thakur, S., & Kumar, R. (2009). Decay studies of 288− 287115 alpha-decay chains. Journal of Physics G: Nuclear and Particle Physics, 36(10), 105104.

[8] Ren, Z., & Xu, C. (2008). Alpha decay half-lives of odd-Z superheavy elements Z= 115→ 113→ 111. In Journal of Physics: Conference Series (Vol. 111, No. 1, p. 012040). IOP Publishing.

[9] Akrawy, D. T., & Poenaru, D. N. (2017). Alpha decay calculations with a new formula. arXiv preprint arXiv:1702.05598.

[10] Akrawy, T. (2017). Theoretical Studies on the α-decay Half-Lives of Even-Even Lv Isotopes. International Journal of Energy and Power Engineering, 6(1), 1-5.

[11] Santhosh, K. P., & Priyanka, B. (2014). Heavy particle radioactivity from superheavy nuclei leading to 298114 daughter nuclei. Nuclear Physics A, 929, 20-37.

[12] Royer, G., & Zhang, H. F. (2008). Recent α decay half-lives and analytic expression predictions including superheavy nuclei. Physical Review C, 77(3), 037602.

[13] Qi, C., Xu, F. R., Liotta, R. J., & Wyss, R. (2009). Universal decay law in charged-particle.

[14] Dong, T., & Ren, Z. (2005). New calculations of α-decay half-lives by the Viola-Seaborg formula. The European Physical Journal A-Hadrons and Nuclei, 26(1), 69-72.

[15] Poenaru, D. N., Plonski, I. H., & Greiner, W. (2006). α-decay half-lives of superheavy nuclei. Physical Review C, 74(1), 014312.

[16] Denisov, V. Y., & Khudenko, A. A. (2009). α-decay half-lives: Empirical relations. Physical Review C, 79(5), 054614.

[17] Krappe, H. J., & Pomorski, K. (2012). Theory of Nuclear Fission: A Textbook (Vol. 838). Springer Science & Business Media.

[18] Poenaru, D. N., Gherghescu, R. A., & Greiner, W. (2011). Single universal curve for cluster radioactivities and α decay. Physical Review C, 83(1), 014601.

[19] Blendowske, R., & Walliser, H. (1988). Systematics of cluster-radioactivity-decay constants as suggested by microscopic calculations. Physical review letters, 61(17), 1930.

[20] Hassanabadi, H., Javadimanesh, E., & Zarrinkamar, S. (2013). A new barrier potential and alpha-decay half-lives of even–even nuclei in the 82≤ Z≤92 regime. Nuclear Physics A, 906, 84-93.

[21] Royer, G., & Moustabchir, R. (2001). Light nucleus emission within a generalized liquid-drop model and quasimolecular shapes. Nuclear Physics A, 683(1), 182-206.

[22] Royer, G. (2010). Analytic expressions for alpha-decay half-lives and potential barriers. Nuclear Physics A, 848(3), 279-291.

[23] Santhosh, K. P., & Priyanka, B. (2014). Predictions for the α-decay chains of Z= 120 superheavy nuclei in the range 272≤ A≤ 319. Physical Review C, 90(5), 054614.

[24] Qi, C., Xu, F. R., Liotta, R. J., Wyss, R., Zhang, M. Y., Asawatangtrakuldee, C., & Hu, D. (2009). Microscopic mechanism of charged-particle radioactivity and generalization of the Geiger-Nuttall law. Physical Review C, 80(4), 044326.

[25] Bao, X. J., Zhang, H. F., Dong, J. M., Li, J. Q., & Zhang, H. F. (2014). Competition between α decay and cluster radioactivity for superheavy nuclei with a universal decay-law formula. Physical Review C, 89(6), 067301.

[26] Santhosh, K. P., Sukumaran, I., & Priyanka, B. (2015). Theoretical studies on the alpha decay of 178–220 Pb isotopes. Nuclear Physics A, 935, 28-42.

[27] Sobiczewski, A., Patyk, Z., & Ćwiok, S. (1989). Deformed superheavy nuclei. Physics Letters B, 224(1-2), 1-4.

[28] Santhosh, K. P., & Priyanka, B. (2014). Predictions for the α-decay chains of Z= 120 superheavy nuclei in the range 272≤ A≤ 319. Physical Review C, 90(5), 054614.

[29] Viola, V. E., & Seaborg, G. T. (1966). Nuclear systematics of the heavy elements—II Lifetimes for alpha, beta and spontaneous fission decay. Journal of Inorganic and Nuclear Chemistry, 28(3), 741-761.

[30] Denisov, V. Y., & Khudenko, A. A. (2010). Erratum: α-decay half-lives: Empirical relations [Phys. Rev. C 79, 054614 (2009)]. Physical Review C, 82(5), 059901.

[31] Denisov, V. Y., & Khudenko, A. A. (2009). α-Decay half-lives, α-capture, and α-nucleus potential. Atomic Data and Nuclear Data Tables, 95(6), 815-835.

[32] Denisov, V. Y., Davidovskaya, O. I., & Sedykh, I. Y. (2015). Improved parametrization of the unified model for α-decay and α capture. Physical Review C, 92(1), 014602.

[33] Audi, G., Kondev, F. G., Wang, M., Pfeiffer, B., Sun, X., Blachot, J., & MacCormick, M. (2012). The NUBASE2012 evaluation of nuclear properties. Chinese Physics C, 36(12), 1157.